\documentclass[10pt,conference]{IEEEtran}
\usepackage{cite}
\usepackage[tight,footnotesize]{subfigure}
\usepackage{graphicx}
\usepackage[cmex10]{amsmath}
\usepackage{amssymb}
\usepackage{mathtools}
\usepackage{multirow}
\usepackage{notoccite}
\usepackage{color} 
\usepackage{epsfig}
\usepackage{latexsym}
\usepackage{subfigure}
\usepackage{pstricks}
\usepackage{pdfpages}
\usepackage{epstopdf}
\DeclareGraphicsExtensions{.eps}
\usepackage{algorithmic}
\usepackage{algorithm}
\usepackage{multirow}
\usepackage{graphicx}
\usepackage{xurl}
\usepackage{bm}

\begin{document}
	\title{Swarm Intelligence Optimization of Multi-RIS Aided MmWave Beamspace MIMO}
	\author{\IEEEauthorblockN{Zaid Abdullah, Mario R. Camana, Abuzar B. M. Adam, Chandan K. Sheemar,\\  Eva Lagunas, and Symeon Chatzinotas \\
    Interdisciplinary Centre for Security, Reliability and Trust (SnT), University of Luxembourg, Luxembourg.\\
   \textit{Corresponding Author}: Zaid Abdullah (Email: zaid.abdullah@uni.lu).}}
	
\maketitle
\begin{abstract}
We investigate the performance of a multiple reconfigurable intelligence surface (RIS)-aided millimeter wave (mmWave) beamspace multiple-input multiple-output (MIMO) system with multiple users (UEs). We focus on a challenging scenario in which the direct links between the base station (BS) and all UEs are blocked, and communication is facilitated only via RISs. The maximum ratio transmission (MRT) is utilized for data precoding, while a low-complexity algorithm based on particle swarm optimization (PSO) is designed to jointly perform beam selection, power allocation, and RIS profile configuration. The proposed optimization approach demonstrates positive trade-offs between the complexity (in terms of running time) and the achievable sum rate. In addition, our results demonstrate that due to the sparsity of beamspace channels, increasing the number of unit cells (UCs) at RISs can lead to higher achievable rates than activating a larger number of beams at the MIMO BS.     

\end{abstract}

\section{Introduction}
Beamspace multiple-input multiple-output (MIMO) refers to systems in which data is multiplexed onto orthogonal spatial beams. One of the main advantages of beamspace MIMO is the high sparsity manifested in the beamspace channel matrix~\cite{sayeed2013beamspace}. Thus, one can select only a small number of available beams without experiencing a large performance degradation in terms of the achievable sum rates~\cite{sayeed2013beamspace, amadori2015low}. 

The fundamental difference between beam selection in beamspace MIMO and antenna selection in conventional MIMO is as follows: In conventional MIMO systems, narrower beams are only possible if more antennas are selected. In contrast, the narrow beam widths in the beamspace MIMO are preserved regardless of the number of selected beams (or radio-frequency (RF) chains). The reason is that in beamspace MIMO, the base station (BS) is equipped with a discrete lens array (DLA) which behaves as a convex lens to direct signals toward different points of the focal surface. As such, it is feasible in beamspace MIMO to achieve significant reductions in hardware complexity and power consumption without sacrificing the high antenna gains. Such properties make beamspace MIMO particularly suitable for communications over higher frequency bands such as millimeter wave (mmWave)~\cite{amadori2015low}. 

Another promising technology for high-frequency communications is the reconfigurable intelligent surface (RIS). An RIS is made up of a large number of tiny
elements known as unit cells (UCs). Each UC manipulates the phase of impinging signals, such that signals reflected by different UCs are added constructively at the receiving node. Therefore, one of the main advantages of RISs is that they can provide virtual communication links when the direct link between two communication nodes is blocked. In addition, the near-passive nature of RISs make them an attractive choice to maintain low levels of hardware complexity and power consumption~\cite{basar2019wireless}. 

To that end, we investigate the performance and optimization of multi-RIS-aided mmWave beamspace MIMO systems. In particular, we design a low complexity optimization scheme to perform joint optimizations with respect to beam selection, power allocation, and RIS profile configuration. In addition, we adopt the low-complexity maximum ratio transmission (MRT) for data precoding. Works on RIS-aided beamspace MIMO have been reported in \cite{wang2021intelligent, elganimi2022irs, alimo2023low}. Unlike our work, the authors in~\cite{wang2021intelligent} considered a single user (UE) scenario, while the authors in~\cite{elganimi2022irs} and \cite{alimo2023low} considered a single RIS scenario and modeled the channels between the RIS and UEs as Rayleigh fading, which is not realistic for practical mmWave systems. 

Our main contributions can be summarized as follows: (i) We propose a beamspace MIMO system with multiple UEs and distributed RISs. The presence of distributed RISs is crucial for providing additional paths leading to enhanced channels rank, and thereby enabling spatial multiplexing, (ii) We design a low-complexity particle swarm optimization (PSO) algorithm to jointly optimize the selected beams, allocated powers, and RIS phase profiles, and (iii) we study the trade-offs in terms of number of selected beams, number of UCs, number of UEs, as well as the complexity of the PSO algorithm on the sum rate performance. 

The rest of this paper is organized as follows: System and channel models are described in Section~\ref{sec: system_model}. In Section~\ref{sec: received signals} we formulate the received signals, sum rate, and the optimization problem. The proposed PSO algorithm is explained in Section~\ref{sec: PSO}. Numerical results are presented in Section~\ref{sec: results}. Finally, concluding remarks are in Section~\ref{sec: Conclusions}. 
\paragraph*{Notations}Matrices and vectors are represented by uppercase and lowercase boldface letters, respectively. The vectors $\mathbf x_c$ and $\mathbf x_{z,c}$ denote the $c$th column of matrices $\mathbf X$ and $\mathbf X_z$, respectively. The $i$th element of the vector $\mathbf x$ is denoted by $\left[\mathbf x \right]_i$ with its absolute value denoted by $\left| \left[\mathbf x \right]_i \right|$, while $\left[\mathbf X \right]_{i,j}$ is the $i$th element of the $j$th column of $\mathbf X$. Assuming $\mathbf x$ is an $N$-dimensional vector, $\left[\mathbf x \right]_{i:j}$ denotes the elements in $\mathbf x$ indexed from $i$ to $j$ ($i < j < N$),  while $\left[\mathbf x \right]_{j:}$ refers to all elements in $\mathbf x$ indexed from $j$ to $N$. For a set $\mathcal S$ containing $M$ integer numbers, its cardinality is denoted by $\left | \mathcal S \right | = M$. In addition, $(\cdot)^*$, $(\cdot)^T$, and $(\cdot)^H$ are the conjugate, transpose, and Hermitian transpose operators, respectively. Finally, $\mathbf I_N$ is the $N\times N$ identity matrix, while $\mathbf X = \text{diag}\{\mathbf x\}$ is a diagonal matrix whose diagonal are the elements of $\mathbf x$. 
\section{System and Channel Models}\label{sec: system_model}
\subsection{System Model}
We consider a time division duplex (TDD) downlink transmission, where a single BS equipped with a DLA of length $L$ communicates with $K$ single-antenna UEs. The direct links between the BS and the UEs are assumed to be blocked, and the communication is facilitated by $J$ RISs, each with $M_j$ UCs forming a uniform linear array (ULA). We denote $M$ as the total number of UCs in all $J$ RISs such that $M = \sum_{j\in\mathcal J} M_j$, where $\mathcal J = \{1, \cdots J\}$ is a set containing the indices of all RISs. Analytically, the DLA at the BS can also be modeled as a ULA with $N$ critically sampled elements with inter-element spacing of $d = \lambda/2$~\cite{sayeed2013beamspace}, where $\lambda$ is the carrier wavelength. This corresponds to a DLA of length $L = \frac{1}{2}\lambda N$~\cite{amadori2015low}.

\begin{figure}
    \centering
    \includegraphics[width=0.61\linewidth]{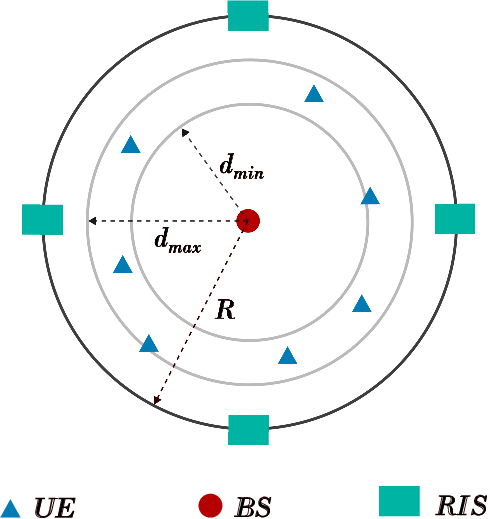}
    \caption{Illustration of the adopted system model with $J = 4$ and $K = 7$.}
    \label{fig:system_model}
    \vspace{-0.2cm}
\end{figure}

We consider a single cell scenario in which the BS is located in the center of the cell which has a radius of $R$ meters, and the $J$ RISs are evenly distributed along the cell circumference. In addition, UEs are randomly located between two circles of radii $d_{min}$ and $d_{max}$ meters, as shown in Fig.~\ref{fig:system_model}.\footnote{In our system, UEs are closer to the cell edge than the BS, which justifies the existence of direct links between UEs and RISs. Also, the BS and RISs can be installed on tall buildings with clear line-of-sight (LoS) links.} 
\subsection{Channel Representation in The Spatial Domain}
MmWave MIMO channels can be characterized in a fashion similar to standard multipath channel models. Hence, the spatial domain channel model is expressed in terms of the corresponding array steering vectors that represent the array phase profile. Specifically, considering a ULA with an arbitrary number of antennas or UC elements $N_x$, one can express the normalized array steering vector for a given angular direction of departing or arriving plane waves as follows~\cite{heath2016overview}:
{\small\begin{equation} 
    \boldsymbol a\pmb{\Big(}\theta, N_x\pmb{\Big)} = \frac{1}{\sqrt{N_x}}\left[1, e^{-\jmath 2\pi\theta}, e^{-\jmath 4\pi\theta}, \cdots, e^{-\jmath 2(N_x-1)\pi\theta} \right]^T,
\end{equation}}where $\theta = \frac{d}{\lambda} \sin(\psi)$ is the spatial frequency, and $\psi$ is the corresponding angle-of-arrival (AoA) or angle-of-departure (AoD). Since the elements are assumed to have half wave-length spacing, the range of values of the spatial frequency is $\theta\in\pmb{[}-0.5, 0.5\pmb{]}$. 
\par The effective cascaded channel in the spatial domain between the BS and all $K$ users, denoted as $\mathbf{\overline{H}} \ \in \mathbb C^{N\times K}$, is:
{\small \begin{equation}
    \mathbf{\overline{H}} = \sum_{j\in\mathcal J} \mathbf{C}_j\boldsymbol{\Phi}_j\mathbf{G}_j,
\end{equation}}where $\mathbf C_j\in \mathbb C^{N\times M_j}$, $\mathbf G_j = \left[\mathbf g_{j,1}, \mathbf g_{j,2}, \cdots, \mathbf g_{j,K}\right] \in \mathbb C^{M_j\times K}$, and $\mathbf \Phi_j = \text{diag}\left\{\left[e^{\jmath [\boldsymbol \phi_j]_{1}}, e^{\jmath [\boldsymbol \phi_j]_2}, \cdots, e^{\jmath [\boldsymbol \phi_j]_{M_j}}\right]\right\}\in \mathbb C^{M_j\times M_j}$ account for the channels between the BS and $j$th RIS, channels between the $j$th RIS and all $K$ UEs, and the diagonal phase-shift matrix at the $j$th RIS, respectively. 
The $k$th column of $\mathbf G_j$, denoted as $\mathbf g_{j,k} \in\mathbb{C}^{M_j\times 1}$, contains the channel coefficients between the $j$th RIS and the $k$th UE. Assuming an LoS and $N_p$ non-LoS (NLoS) paths for all UEs, $\mathbf g_{j,k}$ is expressed as:
{\small \begin{align} \label{chan1}
    \mathbf g_{j,k} & = \mathbf {g}_{j,k,0} + \sum_{l = 1}^{N_p} \mathbf{g}_{j,k,l} \\ \nonumber & = \sqrt{M_j}\eta_{j, k,0} \boldsymbol{a} \pmb{\Big(}\theta_{j,k,0},M_j\pmb{\Big)} + \sqrt{\frac{M_j}{N_p}}\sum_{l = 1}^{N_p}\eta_{j,k,l} \boldsymbol{a}\pmb{\Big(}\theta_{j,k,l},M_j\pmb{\Big)},
\end{align}}where $\mathbf{g}_{j,k,0}$ is the LoS channel vector with $\eta_{j,k,0}$ and $\theta_{j,k,0} = \frac{d}{\lambda}\sin (\psi_{j,k,0})$ being the LoS path gain and direction, respectively. Similarly, $\mathbf{g}_{j,k,l}$ reflects the $l$th NLoS path, with $\eta_{j,k,l}$ and $\theta_{j,k,l} = \frac{d}{\lambda}\sin (\psi_{j,k,l})$ respectively account for the $l$th NLoS path gain and direction. Further, the channel between the BS and the $j$th RIS, denoted as $\mathbf C_j$, is:
{\small \begin{align} \label{chan2}
    \mathbf C_{j} & = \mathbf {C}_{j,0} + \sum_{l = 1}^{N_p} \mathbf{C}_{j,l} \nonumber \\ & = \sqrt{M_jN}\eta_{j,0} \boldsymbol{a} \pmb{\Big(}\theta_{j,0},N\pmb{\Big)} \boldsymbol{a} \pmb{\Big(}\bar{\theta}_{j,0},M_j\pmb{\Big)}^H  \nonumber \\ & + \sqrt{\frac{M_jN}{N_p}}\sum_{l = 1}^{N_p}\eta_{j,l} \boldsymbol{a}\pmb{\Big(}\theta_{j,l},N\pmb{\Big)}\boldsymbol{a}\pmb{\Big(}\bar{\theta}_{j,l},M_j\pmb{\Big)}^H,
\end{align}}where $\theta_{j,i} = \frac{d}{\lambda}\sin (\psi_{j,i})$ and $\bar{\theta}_{j,i} = \frac{d}{\lambda}\sin (\bar{\psi}_{j,i})$ correspond to the angles of departure and arrival, respectively, for the $i$th path with $i \in \{0, \cdots, N_p\}$.
\subsection{Channel Representation in The Beamspace Domain}
The spatial domain channel matrix $\mathbf {\overline H}$ can be transformed into the beamspace domain via the beamforming matrix $\mathbf U$, which reflects the operation of a perfectly designed DLA~\cite{amadori2015low}. Specifically, the $N$ columns in $\mathbf U \in \mathbb C^{N\times N}$ are steering vectors for $N$ fixed spatial angles having a uniform spacing of $\Delta\theta_0 = \frac{1}{N}$, such that the $n$th column of $\mathbf U$ is expressed as:
{\small \begin{equation} 
    \mathbf u_n = \boldsymbol{a} \pmb{\Big(}\tilde n\Delta \theta_0,N\pmb{\Big)},
\end{equation}}where $\tilde n = (n-1) - 0.5(N-1)$. The matrix $\mathbf U$ resembles a Discrete Fourier Transform (DFT) operation, and it is a unitary matrix satisfying $\mathbf U \mathbf U^H = \mathbf U^H\mathbf U = \mathbf I_N$.
The beamspace channel matrix $\mathbf H\ \in \mathbb C^{N\times K}$ is obtained as:
{\small \begin{equation} \label{beam channel}
    \mathbf H = \mathbf U \mathbf{\overline{H}} = \mathbf U \sum_{j\in\mathcal J} \mathbf C_j \mathbf \Phi_j \mathbf G_j.
\end{equation}}Since $\mathbf U$ is unitary, the beamspace channel matrix $\mathbf H$ is an equivalent representation of the spatial channel matrix $\mathbf {\overline{H}}$~\cite{sayeed2013beamspace}. 

Next, we formulate the received signals, signal-to-interference-and-noise ratios (SINRs), the achievable sum rate, as well as the considered optimization problem.
\vspace{-0cm}
\section{Received Signals, SINRs and Problem Formulation} \label{sec: received signals}
\vspace{-0cm}
\subsection{Received Signals}
The deployment of a DLA at the BS means that performing signal processing and expressing the received signals can be achieved in the beamspace domain. As such, assuming all $N$ beams are selected, the received signal at the $k$th UE is:
\small{\begin{equation}\label{eq: y_k}
    y_k = \sqrt{p_k} \mathbf h^T_k \mathbf w_k x_k + \sum_{i \in \mathcal K\setminus k} \sqrt{p_i} \mathbf h_k^T \mathbf w_i x_i + n_k,
\end{equation}}where $\mathcal K = \{1, 2, \cdots, K\}$ is a set containing the indices of all UEs. Focusing on the $k$th UE, $x_k$ is the information symbol intended for the UE satisfying $\mathbb E\{|x_k|^2\} = 1$, $p_k$ is the amount of power allocated for the UE, $\mathbf w_k = \frac{\mathbf h_k^*}{\left\| \mathbf h_k\right\|}$ is the unit-norm MRT precoding vector, and $n_k$ is the additive white Gaussian noise with zero mean and variance of $\sigma^2$. Regarding the second term in the right hand side of (\ref{eq: y_k}), it accounts for the inter-user interference.\footnote{We assume a perfect knowledge of channel state information (CSI).}
\vspace{-0cm}
\subsection{SINRs and Achievable Rates }
The received SINR at the $k$th UE is given as:
\small{\begin{equation} \label{SINRk}
    \mathrm{SINR}_k = \frac{p_k \left|\mathbf h_k^T\mathbf w_k \right|^2}{\sum_{i\in \mathcal K\setminus k}p_i \left| \mathbf h_k^T\mathbf w_i \right|^2 + \sigma^2},
\end{equation}}and the achievable rate for the same UE (in bit/s/Hz) is:
\small{\begin{equation} 
    \mathrm R_k = \log_2\left(1 + \mathrm{SINR}_k\right).
\end{equation}}Then, the total achievable sum rate can be expressed as follows:
{\small \begin{equation} \label{sum rate}
    \mathcal R = \sum_{k\in\mathcal K} \mathrm R_k.
\end{equation}}
\subsection{Problem Formulation}
Our aim is to maximize the achievable sum rate via selecting $N_s \ge K$ out of $N$ beams,\footnote{The number of required RF chains at the BS is equivalent to the number of selected beams. Since there are $K$ UEs, $N_s$ cannot be smaller than $K$.} optimizing the phase profile of the $J$ RISs, and optimizing the power distribution for the $K$ UEs. The corresponding optimization problem is given as:
{\small \begin{align} \label{OP1}
	& \hspace{.1cm}\underset{\substack{\mathbf p, \mathbf \Delta, \mathbf \Phi }}{\text{maximize}} \hspace{1cm}  \mathcal R\left(\mathbf p, \mathbf \Phi, \mathbf \Delta\right) \\ 
	&\hspace{.1cm}\text{subject to} \nonumber \\ 
	& \hspace{.1cm} \sum_{k \in \mathcal K} p_k \le P, \label{1st_const} \\ & 
  \hspace{.1cm} \left[\mathbf \Delta\right]_{n,n} \in \{0, 1\}, \forall n \in \mathcal N \label{2nd_const} \\ & \sum_{n\in\mathcal N} \left[ \mathbf \Delta \right]_{n,n} = N_s, \label{3rd_const} \\ & \hspace{.1cm} \left|\left[\mathbf \Phi_j\right]_{m,m}\right| = 1, \forall j \in \mathcal J, \forall m \in \mathcal M_j \label{4th_const}
\end{align}}where the objective function $\mathcal R\left(\mathbf p, \mathbf \Phi, \mathbf \Delta\right)$ is given in (\ref{eq: rate_opt}) at the top of the next page, $\mathbf p = \{p_1, \cdots, p_K\}$, $\mathbf \Phi = \{\mathbf \Phi_1, \cdots, \mathbf \Phi_J\}$, $\mathcal N = \{1, \cdots, N\}$, and $\mathcal M_j = \{1, \cdots, M_j\}$. Constraint (\ref{1st_const}) ensures that the total available transmit power at the BS $P$ is not exceeded. The variable $\mathbf \Delta$ is a diagonal matrix that controls which beams are selected. Specifically, in constraint (\ref{2nd_const}), if the $n$th beam is selected, then $[\mathbf \Delta]_{n,n}$ will be equal to $1$, otherwise $[\mathbf \Delta]_{n,n}=0$. In addition, constraint (\ref{3rd_const}) ensures that exactly $N_s$ out of the available $N$ beams are selected. Finally, constraint (\ref{4th_const}) ensures that the unit-modulus constraint at the RIS is not violated. 
\begin{figure*}[t!]
    \small {\begin{equation} \label{eq: rate_opt}
        \mathcal R\left(\mathbf p, \mathbf \Phi, \mathbf \Delta\right) = \sum_{k\in\mathcal K} \log_2\left(1 + \frac{p_k\left| \left( \mathbf U \sum_{j \in \mathcal J} \mathbf C_j \mathbf \Phi_j 
\mathbf g_{j,k} \right)^T \mathbf \Delta \ \frac{\left(\mathbf U \sum_{j \in \mathcal J} \mathbf C_j \mathbf \Phi_j \mathbf g_{j,k}\right)^*}
{\left\| \mathbf \Delta \mathbf U \sum_{j \in \mathcal J} \mathbf C_j \mathbf \Phi_j 
\mathbf g_{j,k} \right\|} \right|^2 }{\sum_{i\in\mathcal K\setminus k} p_i \left| \left( \mathbf U \sum_{j \in \mathcal J} \mathbf C_j \mathbf \Phi_j 
\mathbf g_{j,k} \right)^T \mathbf \Delta \ \frac{\left(\mathbf U \sum_{j \in \mathcal J} \mathbf C_j \mathbf \Phi_j \mathbf g_{j,i}\right)^*}
{\left\| \mathbf \Delta \mathbf U \sum_{j \in \mathcal J} \mathbf C_j \mathbf \Phi_j 
\mathbf g_{j,i} \right\|} \right|^2 + \sigma^2 }\right).
    \end{equation}}
    \hrule 
\end{figure*}
The optimization problem in (\ref{OP1}) is highly non-convex due to the unit-modulus constraint, the inter-user interference terms in the objective function, as well as the binary beam selection matrix $\mathbf \Delta$, which makes finding the optimal solution an NP-hard problem. To tackle these challenges, we design a low-complexity Metaheuristic approach and perform joint optimization of all optimization variables as described in detail in the next section. 
\section{Particle Swarm Optimization Approach} \label{sec: PSO}
In this section, we design a low-complexity PSO algorithm to jointly solve the optimization problem in (\ref{OP1}). The PSO is a Metaheuristic algorithm that makes few or no assumptions about the problem being optimized. It consists of a population, known as a swarm, of possible solutions (also known as particles). The movement of particles in the search space is influenced by the positions of their local neighbors, as well as the position of the best particle in the swarm, that is, the particle that achieves the best result~\cite{kennedy1995particle}. 

In the following subsections, we explain the steps of the proposed PSO algorithm. 
\subsection{Initialization}
The first step is to randomly initialize the population matrix $\mathbf F \in \mathbb R^{N_v\times A}$, where $N_v = N + K + M$ is the total number of optimization variables, and $A$ is the number of particles in the population, each representing a possible solution. Considering the $a$th column in $\mathbf F$, denoted as $\mathbf f_a \in \mathbb R^{N_v\times 1}$ with $a\in \{1, \cdots, A\}$, the first $N$ elements in $\mathbf f_a$, denoted as $\left[\mathbf f_a\right]_{1:N}$ are associated with the beam selection subproblem. In addition, the elements $\left[\mathbf f_a\right]_{N+1: N+K}$ are associated with the power allocation subproblem, while the last $M$ elements in $\mathbf f_a$, i.e. $\left[\mathbf f_a\right]_{N+K+1:}$, are associated with the phase values of all $M$ UCs. 

In addition, the velocity matrix $\mathbf X \in\mathbb R^{N_v\times A}$, which controls the movement and the speed of particles, is initialized with zeros, and will be utilized to update the population as will be seen in subsection~\ref{D}. 
\subsection{Constraints Check} \label{B}
Once the population is initialized or updated, its elements must be checked to ensure that no constraints are violated. In the following, we explain how this is done for each subproblem. While our focus in this subsection will be on fixing the entries of a single particle $\mathbf f_a$, the following steps are applied for all $A$ particles in the population. 
\subsubsection{Beam Selection Constraint}For each column in $\mathbf F$, the following operation is performed:
{\small \begin{equation} \label{eq: beam norm}
    \left[\mathbf f_a\right]_{n} := \frac{\left|\left[\mathbf f_a\right]_n\right|}{\max\left\{\left|\left[\mathbf f_a\right]_{1}\right|, \cdots, \left|\left[\mathbf f_a\right]_{N}\right| \right\}}, \ \ \ \forall n \in \{1, \cdots, N\}.
\end{equation}}The normalization in (\ref{eq: beam norm}) ensures that for each particle, all $N$ entries corresponding to the beam selection subproblem are positive and within the range $[0,1]$.
\subsubsection{Power Allocation Constraint}The entries corresponding to the power allocation subproblem are normalized as follows:
{\small \begin{equation} \label{eq: power norm}
    \left[\mathbf f_a\right]_{n} := \frac{\left|\left[\mathbf f_a\right]_n\right|}{\sum_{i = N+1}^{N+K}\left|\left[\mathbf f_a\right]_i\right|} \times P, \ \ \ \forall n \in \{N+1, \cdots, N+K\}.
\end{equation}}In (\ref{eq: power norm}), we ensure that the amount of power allocated for each UE is positive, and the total transmit power for all UEs is equal to $P$.
\subsubsection{RIS Profile Constraint}The final step is to adjust the phase-shift values of all $M$ UCs to be within the range $\left[0, 2\pi\right]$ as follows:
{\small \begin{equation} \label{adjust}
	\left[ \mathbf f_a \right]_{n} := \begin{cases}
		\left[ \mathbf f_a \right]_n + 2\pi, \ \ \ \ \ \  \text{if} \left[ \mathbf f_a \right]_n < 0 \\ \left[ \mathbf f_a \right]_n - 2\pi, \ \ \ \ \ \ \text{if} \left[ \mathbf f_a \right]_n > 2\pi,
	\end{cases}
\end{equation}}$\forall n \in\{N+K+1, \cdots, N_v\}$.
\subsection{Evaluate the Quality of Particles}
After we ensure that all constraints are met, we evaluate the quality, which in this case is the achievable sum rate in (\ref{eq: rate_opt}), of each particle. Focusing on an arbitrary particle $\mathbf f_a$, the evaluation involves performing the following steps:
\subsubsection{Construct the RISs Profile Matrices}The first step is to construct the response of the $J$ RISs with the optimized phase shifts. We denote $\boldsymbol \Phi(a) = \{\boldsymbol \Phi_1(a), \cdots, \boldsymbol \Phi_J(a)\}$ as the RISs profile matrices where the response of all RISs are obtained from $\mathbf f_a$ as:
{\small \begin{equation} \label{eq: chan}
    \boldsymbol \Phi_j(a) = \left\{\boldsymbol \Phi_j \middle| [\boldsymbol \phi_j]_1, \cdots [\boldsymbol \phi_j]_{M_j} = \left[ \mathbf f_a\right]_{z_j+1: z_j + M_j} \right\},
\end{equation}}where $z_j = N + K + \sum_{i = 1}^{j-1} M_i$, and $j\in \mathcal J$.
\subsubsection{Selection of $N_s$ Beams} For each particle, we select the $N_s$ beams with the highest values in $[\mathbf f_a]_{1:N}$ as follows:
{\small \begin{equation}
    \mathcal S(a) = \arg \underset{N_s}{\max} \left\{ [\mathbf f_a]_{1}, \cdots, [\mathbf f_a]_{N}\right\},
\end{equation}}where $\mathcal S(a)$ is the set of indices of selected beams for particle $\mathbf f_a$, with cardinality $\left|\mathcal S(a)\right| = N_s$. Then, the diagonal elements in $\boldsymbol \Delta(a)$ that correspond to the $N_s$ values in $\mathcal S(a)$ are set to $1$, while the remaining $N - N_s$ diagonal elements in $\boldsymbol \Delta(a)$ are set to $0$. It should be noted that $\boldsymbol \Delta(a)$ denotes the diagonal binary beam selection matrix $\boldsymbol \Delta$ for the $a$th particle. Analytically, the relation between $\boldsymbol \Delta(a)$ and $\mathcal S(a)$ can be expressed as:
{\small \begin{equation}
    \left[ \boldsymbol \Delta(a) \right]_{n,n} = \begin{cases}
		1, \ \ \ \ \ \  \text{if} \ \ n \in \mathcal S(a) \\ 0, \ \ \ \ \ \ \text{otherwise},
	\end{cases}, \forall n \in \mathcal N.
\end{equation}}
\subsubsection{Power Allocation}
The final step before evaluating the quality is to obtain the power allocated for each UE as follows:
{\small \begin{equation}
    p_k (a) = \left[\mathbf f_a\right]_{N+k}, \ \ \ \forall k \in \{1, \cdots, K\}.
\end{equation}}For the $a$th particle, we denote the set of power values for all UEs as $\mathbf p(a) = \{p_1(a), \cdots, p_K(a)\}$. 
\subsubsection{Sum Rate Evaluation}Now we can evaluate the quality (i.e. the achievable sum rate) of the $a$th particle by substituting $\mathbf p = \mathbf p(a)$, $\boldsymbol \Phi = \boldsymbol \Phi(a)$, and $\boldsymbol \Delta = \boldsymbol \Delta(a)$ in (\ref{eq: rate_opt}).
\subsection{Find Local and Global Best Solutions}
Once the qualities of all $A$ particles are evaluated, we find the local and global best solutions. Specifically, the global best solution is the particle in $\mathbf F$ that achieves the highest sum rate. We denote the global best solution as $\mathbf f_{gb}$. 

On the other hand, the local best solution for the $a$th particle is the particle that achieves the highest sum rate between only the two neighboring particles of $\mathbf f_a$, which are $\mathbf f_{a+1}$ and $\mathbf f_{a-1}$.\footnote{We adopt a ring topology for the particles. Hence, the neighboring particles of $\mathbf f_1$ are $\mathbf f_2$ and $\mathbf f_A$, while the neighbors of $\mathbf f_A$ are $\mathbf f_1$ and $\mathbf f_{A-1}$.} We denote the local best solution for the $a$th particle as $\mathbf f_{lb,a}$.
\subsection{Update the Population} \label{D} The final step is to update the population. This is performed over two stages. At first, the velocity of the population $\mathbf X$ is updated as:
{\small \begin{align}
    \left[\mathbf X\right]_{i,a} := & \mu \left[\mathbf X\right]_{i,a} + w_1 * \textit{rand}_1 \left(\left[\mathbf f_{gb}\right]_i - \left[\mathbf f_a\right]_i\right) \nonumber \\ +\  & w_2 * \textit{rand}_2 \left(\left[\mathbf f_{lb,a}\right]_i - \left[\mathbf f_a\right]_i\right),
\end{align}}where $a\in \{1, \cdots, A\}$, $i\in\{1, \cdots, N_v\}$, $\textit{rand}_1$ and $\textit{rand}_2$ are random numbers between $0$ and $1$ drawn from a uniform distribution, $w_1$ and $w_2$ are learning factors, and $\mu$ is the inertia weight. 

Next, the population matrix is updated as:
{\small \begin{equation}
    \mathbf F := \mathbf F + \mathbf X.
\end{equation}}
\subsection{Iterate}The operations explained in \ref{B} to \ref{D} are for a single iteration of the PSO algorithm. Therefore, the same steps \ref{B} to \ref{D} will be repeated $T$ times, where $T$ is the total number of optimization iterations. At the end, the particle that achieves the highest sum rate across all $T$ iterations will be selected as the final solution. 

It is worth mentioning that the proposed PSO approach has a per-iteration complexity order of $\mathcal O\left(A\cdot N\cdot M\cdot K\right)$.  
\section{Simulation Results} \label{sec: results}
\subsection{Channel Parameters and Path-Loss Model}
We adopt the Urban Micro (UMi) path-loss (PL) model in~\cite[Table 7.4.1-1]{3GPP_Channel}. For any two arbitrary nodes $i$ and $j$, the corresponding PL, denoted as $\zeta_{i,j}$, can be expressed as
{\small \begin{equation}
\zeta_{i,j}\ [\mathrm{dB}] = \begin{cases}
 32.4 + 21\log_{10}(d_{i,j}) + 20\log_{10}(f_c), \hspace{1.17cm} \text{if}\  \text{LoS} \\ 32.4 + 31.9\log_{10}(d_{i,j}) + 20\log_{10}(f_c), \hspace{0.7cm} \text{if}\  \text{NLoS}
\end{cases}
\end{equation}}where $f_c$ is the operating frequency in GHz, and $d_{i,j}$ reflects the distance in meters. The values of $\eta$ in (\ref{chan1}) and (\ref{chan2}) are then obtained as $\eta_{i,j,0} = \sqrt{10^{\frac{-\zeta_{i,j}}{10}}}$ for the LoS, and $\eta_{i,j,l} = \sqrt{10^{\frac{-\zeta_{i,j}}{10}}} e^{-\jmath 2\pi \vartheta_l}$ for the NLoS, where $\vartheta_l$ is uniformly distributed between $0$ and $2\pi$. In addition, the AoAs at RISs are drawn from a uniform distribution between $-\pi/2$ and $\pi/2$, while the AoDs from the BS as well as the AoAs at all UEs are uniformly distributed between $-\pi$ and $\pi$.
\subsection{Simulation Parameters}
Unless stated otherwise, the following set of parameters where adopted in the simulations: The carrier frequency $f_c = 30$ GHz, the noise variance $\sigma^2 = -110$ dBm, number of RISs $J = 8$, number of UEs $K = 8$, cell radius $R = 40$ meters with $d_{min} = 25$ meters and $d_{max} = 35$ meters, number of NLoS paths $N_p = 2$, number of DLA elements $N = 64$, total transmit power $P = 40$ dBm, the number of selected beams $N_s = 8$, and $M_j = M/J$ $\forall j\in\mathcal J$. In addition, for the PSO algorithm, the number of particles $A = 50$, the number of iterations $T = 200$, the inertia weight $\mu = 0.05$, and the PSO learning parameters $w_1 = w_2 = 2$. Simulations are averaged over $1000$ independent channel realizations and UEs locations.

\begin{figure}[t]
    \centering
    \includegraphics[width=7.5cm,height=5.8cm]{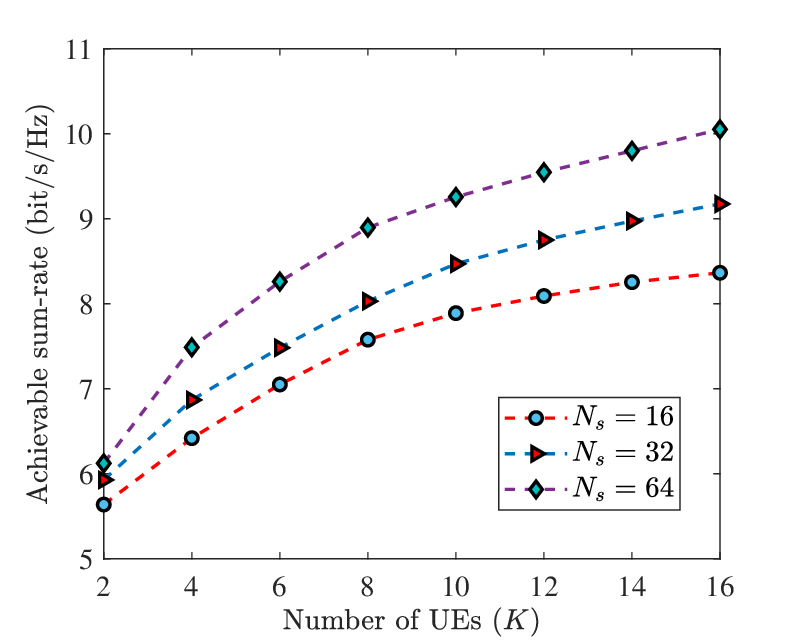}
    \caption{Achievable sum-rate vs. number of UEs for different number of selected beams when $M = 128$.}
    \label{fig:UEs vs beams}
\end{figure}
\subsection{Discussions}
Fig.~\ref{fig:UEs vs beams} shows the sum rate for different numbers of UEs and selected beams. Accommodating a larger number of UEs leads to higher achievable sum rates. In addition, the results reflect the key advantage of the beamspace MIMO which relies on high channel sparsity. In particular, as the number of selected beams increases, a relatively small increase in the total sum rate is observed. For example, when  $K = 16$, the sum rate increases by only $0.8$ bits/s/Hz when the number of $N_s$ is doubled from $16$ to $32$.

\begin{figure}
    \centering
    \includegraphics[width=7.5cm,height=5.8cm]{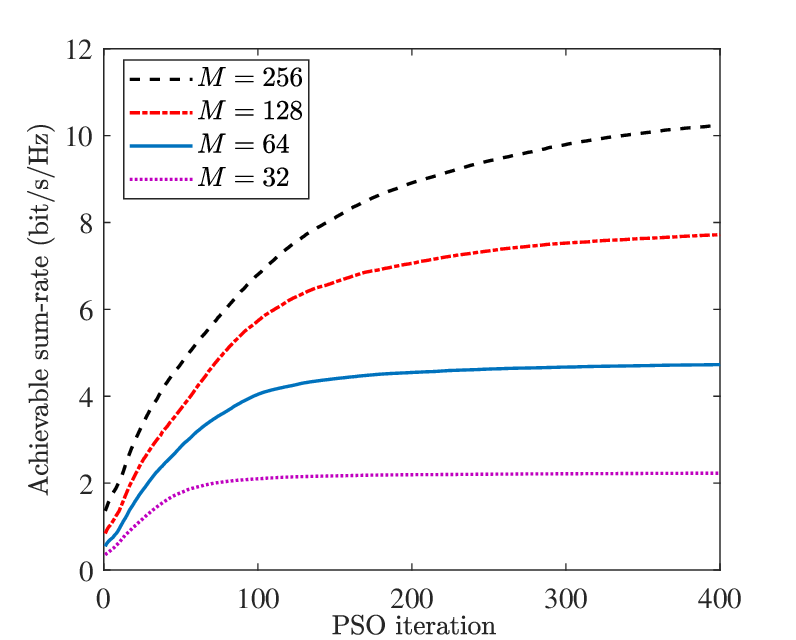}
    \caption{Convergence of the proposed PSO scheme, $K = N_s = 8$.}
    \label{fig:convergence}
\end{figure}

The efficiency of the proposed PSO algorithm can be seen in Fig.~\ref{fig:convergence}, which shows the achievable sum rate as a function of the number of PSO iterations. It should be noted that when the number of iterations is equal to zero, the results reflect the achievable sum rate with random solutions of selected beams, allocated powers, and phase-shifts of UCs. For the case where $M = 256$, the proposed PSO scheme provides more than $7$ bit/s/Hz improvement compared to the random solution after $200$ optimization iterations, and more than $8$ bit/s/Hz after $400$ iterations. Furthermore, it can also be observed that increasing the number of optimization variables (in this case $M$) affects the convergence behavior of the PSO algorithm. Specifically, the PSO algorithm requires a larger number of iterations to converge when the number of optimization variables increases. This behavior is natural due to the larger search space the PSO needs to explore when increasing the number of optimization variables. 

Finally, we observe from Fig.~\ref{fig:UEs vs beams} and Fig.~\ref{fig:convergence} that increasing the number of UCs is preferred to increasing the number of selected beams. For example, increasing $N_s$ from $32$ to $64$ corresponds to an increase of $0.9$ bit/s/Hz in the sum rate (or even less, depending on the number of UEs). In contrast, increasing $M$ from $32$ to $64$ can lead to a sum rate that is $2$ bit/s/Hz higher. This is due to the sparsity of the beamspace channels. More specifically, increasing the number of beams will result in additional channel vectors that are likely not well aligned with the directions of RISs, and thus have low channel gain. In contrast, increasing the number of UCs $M$ can enhance the channel gain for beams pointing toward the directions of RISs. This brings significant advantages in terms of energy efficiency, as increasing the number of beams means more power-hungry RF chains at the BS, while incorporating additional UCs at RISs does not require additional RF chains. 

\section{Concluding Remarks}\label{sec: Conclusions}
The performance of a multi-RIS-aided mmWave beamspace MIMO with multiple UEs was investigated in this work. A low-complexity algorithm based on the PSO approach was designed to jointly optimize the selection of beams, allocation of powers, and configuration of the multiple RISs. Numerical results demonstrated the efficiency of the proposed optimization approach, and highlighted key complexity-performance trade-offs. It was also demonstrated that with beamspace MIMO, one should aim for larger RISs rather than activating more beams at the MIMO BS, which could result in significant gains in terms of the energy efficiency performance. 
\section*{Acknowledgment}
This work has been supported by the Smart Networks and Services Joint Undertaking (SNS JU) project TERRAMETA under the European Union’s Horizon Europe research and innovation programme under Grant Agreement No 101097101, including top-up funding by UK Research and Innovation (UKRI) under the UK government’s Horizon Europe funding guarantee.

\bibliographystyle{IEEEtran}
\bibliography{BeamSpace_MIMO}	
\end{document}